\documentclass[twocolumn,jcp]{revtex4-1}
\usepackage{graphicx}
\usepackage{rotating}
\usepackage{amsmath}
\usepackage{amsfonts}
\usepackage{amssymb}
\usepackage{enumerate}
\usepackage{longtable}
\usepackage{dcolumn}
\usepackage{bm}
\usepackage{latexsym}
\usepackage{epsfig}
\usepackage{subfigure}
\newcommand{\be}{\begin{equation}}
\newcommand{\ee}{\end{equation}}
\newcommand{\bea}{\begin{eqnarray}}
\newcommand{\eea}{\end{eqnarray}}

\newcommand\f{\frac}

\begin{document}

\title{Pressure Driven Phase Transition in 1T-TiSe$_2$, a MOIPT+DMFT Study}
\author{S. Koley}
\affiliation{St. Anthony's College, Shillong, Meghalaya, 793001, India}
\begin{abstract}
The nature of unconventional superconductivity associated with charge density 
wave order in transition metal dichalcogenides is currently a debated issue. 
Starting from a normal state electronic structure followed by a charge ordered 
state how superconductivity 
 in 1T-TiSe$_2$ arises with applied pressure is still under research. 
A preformed excitonic liquid driven ordered state 
mediated superconductivity is found in broad class of TMD on the 
border of CDW. Using dynamical mean 
field theory with input from noninteracting band structure calculation, 
I show a superconducting phase appears near about 2 GPa 
pressure at a temperature of 2 K and this region persists upto 4 GPa Pressure.
\end{abstract}
\maketitle

\section{Introduction}

Despite its advanced age, superconductivity is still one of the hottest topic 
in the strongly correlated materials \cite{hosno,AL,AS,Anderson}. It has been considered the most 
extraordinary and mysterious property of materials for a long time. 
High-T$_c$ superconductors might look like an evergreen research theme \cite{pwa}. 
The main reason behind this is the number of opened 
questions concerning the pairing mechanism which is strongly related to the 
materials' electronic structure \cite{pwa}. The basic and till date most comprehensive 
theory of superconductivity was introduced by Bardeen, Cooper and Schrieffer~\cite{bcs}.
Bardeen et al., discovered that an existing attractive force between two 
electrons makes energetically favorable bound two electron state and 
they can act as boson to condensate without violation of Pauli exclusion 
principle.\\
 
\noindent The low dimension and other structural symmetry make the transition 
metal dichalcogenide (TMD) system likely to be charge density wave (CDW) ordered. 
Energy minimization of these type of electronic system leads to 
the CDW transition.
In the CDW ordered phase the structural periodicity of the system is reorganized
 to achieve stable state, which affects conductivity of the material \cite{aipp,attise2,arghya}. 
Ordering in a CDW state and superconducting state are structurally two 
different phenomena but their 
coexistence and competitive nature is found in many systems \cite{kuma,nunez}.
Condensed matter physics has a lots of findings both theoretically and 
experimentally on these two competing ground states \cite{chang,gabovich}. The findings like 
Fermi surface instabilities, electron-phonon coupling, 
orbital selectivity and antiferromagnetic ground state
 lead to both CDW and superconducting phase \cite{scalapino,valla,attise2,chang}.
TiSe$_2$ was one of the first known compounds with CDW ground state, and 
also most studied material till now due to the puzzling nature of its 
CDW transition. Here the CDW 
transition at a temperature 200 K is to a commensurate state with a 
$2\times 2 \times 2$ wavevector
\cite{disalvo,woo}. Most of the findings in TiSe$_2$ points that the idea of 
Fermi surface nesting as a reason for CDW transition is not applicable here.
The normal state of this material is explained as a small indirect band gap 
semimetal \cite{disalvo,wilson2,zunger,stoffel}. 
Latest experiments on TiSe$_2$ like 
angle resolved photoemission spectra~\cite{kidd} explained that 
the CDW phase in TiSe$_2$ consists of larger indirect band gap at different 
momentum direction of brillouin zone. Theoretically CDW here is 
predicted by electron-phonon coupling and exciton induced orbital selectivity
~\cite{attise2}, though the CDW ground state and low temperature 
superconductivity still remains controversial. 

\noindent Recent finding of superconductivity (SC) in doped 1T-TiSe$_2$ 
prompted large number of intense research activities~
\cite{morosan,zhao,li,cerce,qian} 
due to the questionable SC transition following charge density wave (CDW)~
\cite{wilson,wilson1}. The similarity in the pressure temperature 
phase diagram of 1T-TiSe$_2$ with  
other strongly correlated materials and the semi-metallic behavior of the 
compound in normal and density wave 
led to the idea of Overhauser type CDW transition~\cite{over}, 
i.e. Bose-Einstein condensation of correlated excitons. 
The superconductivity in Cu$_x$TiSe$_2$ was 
found after the CDW phase due to doping of Cu confined in a region around 
critical doping~\cite{morosan}. Corresponding 
transition temperatures form a dome-like structure with copper doping. 
Moreover dome formation is mostly found as the important charatestics 
of phase diagrams found in heavy fermion compounds, cuprate superconductors 
and layered materials~\cite{lee, 
sidorov}. However superconductivity in these high temperature superconductors 
 is explained as a competing order with antiferromagnetic ground state
\cite{sidorov,tallon} whereas in 1T-TiSe$_2$ 
there is a novel type of superconductivity, where emergence of 
SC ordering 
is not related to any magnetic degrees of freedom~\cite{li,barath}.
Alternatively the SC state close to the CDW state  
of the parent compound has no effect on each other   
and the superconductivity here is predicted as a conventional 
 SC phase driven by phonon~\cite{zhao}. Around the transition temperature 
the dome formation with 
doping is then explained as a consequence of 
the shifting of chemical potential above Fermi level caused by the 
doping induced electrons~\cite{zhao}.
Though these issues have been explained previously with different 
theoretical views 
but there is lack of any consistent picture how CDW state evolve into 
superconducting state with pressure. Main goal of this paper is to find out 
a consistent theory describing CDW state to superconducting state transition.
 What is behind the normal to density wave ordering in 1T-TiSe$_2$ has 
been already explained \cite{attise2} from a preformed excitonic liquid (PEL) view which 
causes orbital selectivity 
and CDW phase. This PEL view \cite{arghya} 
is now proved to be a novel alternative to conventional theories for TMD. 
I will start from this PEL view to find out high pressure phase 
at high temperature 
and then I have added a symmetry breaking term to the Hamiltonian to get 
low temperature ordered states. 

\section{Method}
\noindent For 1T-TiSe$_{2}$, normal state followed by incoherent CDW state with two 
particle instability is explained from LCAO + dynamical-mean field theory (DMFT)
 calculation of a two band extended Hubbard model \cite{attise2}. The two band model I 
used to find normal state is 
$$H=\sum_{{\bf k},l,m,\sigma}(t_{\bf k}^{lm}+\mu_{l}\delta_{lm})c_{{\bf k}l\sigma}^{\dag}c_{{\bf k}m\sigma}+ U\sum_{i,l=a,b}n_{il\uparrow}n_{il\downarrow}+ $$
$$U_{ab}\sum_{i}n_{ia}n_{ib}+ g\sum_{i}(A_{i}+A_{i}^{\dag})(c_{ia}^{\dag}c_{ib}+h.c)$$ 
$$+\omega_0\sum_{i}A^{\dagger}_{i}A_{i}$$\\ 
It has been observed that in the normal 
state of 1T-TiSe$_{2}$ one-particle inter-band hopping is ineffective
 so the ordered states must now arise directly as two-particle instabilities.
The two particle interaction, obtained to second order is proportional to 
$t_{ab}^2$, which is more relevant in ordered low $T$ region. The interaction 
is $H_{res}\simeq -t_{ab}^{2}\chi_{ab}(0,0)\sum_{<i,j>,\sigma\sigma'}
c_{ia\sigma}^{\dag}c_{jb\sigma}c_{jb\sigma'}^{\dag}c_{ia\sigma'}$,
with $\chi_{ab}(0,0)$ the inter-orbital susceptibility
calculated from the normal state DMFT results. 
Now the new effective Hamiltonian is $H=H_{n}+H_{res}^{HF}$, where
$H_{n} = \sum_{k,\nu}(\epsilon_{k,\nu}+\Sigma_{\nu}(\omega)-E_{\nu})
c_{k,\nu}^{\dagger}
c_{k,\nu}+\sum_{a\ne b,(k)}t_{ab}( c_{k,a}^{\dagger}c_{k,b}+h.c.)$,
with $\nu=a,b$.
The residual Hamiltonian $H_{res}^{HF}$ is found by 
decoupling the intersite interaction in a generalised HF sense. Now this will 
produce two 
competing instabilities, one in particle-hole channel representing a CDW and 
another in particle-particle (SC) channel. The HF Hamiltonian is  
$H_{res}^{HF}=
-p\sum_{\langle i,j\rangle,a,b,\sigma,\sigma'}(\langle n_{i,a}\rangle
n_{j,b}+\langle n_{j,b}\rangle n_{i,a}-\langle
c_{i,a,\sigma}^{\dagger}c_{j,b,\sigma'}^{\dagger}\rangle c_{j,b,\sigma'}c_{i,a,\sigma}+ h.c.)$ (where p is proportionality constant). The CDW phase 
can be found directly from the particle hole instability with the order 
parameter $\Delta_{CDW} \propto \langle n_{a}-n_{b} \rangle $. I studied 
here the superconducting phase with the two particle instability in 
particle-particle channel (with parametrized p=0.1) and the superconducting order parameter 
can be calculated from $\Delta_{sc} \propto \langle c_{i,a,\sigma}^{\dagger}c_{j,b,\sigma'}^{\dagger}\rangle$ which yields multiband spin-singlet SC.\\
 
The input for DMFT is taken from the pressure dependent LDA DOS which
can be calculated from bandwidth pressure relation $D=D_0exp(\Delta p/K)$ 
(where 
$D$ is bandwidth, $D_0$ is set 1 for reference value, K is bulk modulus and 
$\Delta p$ is change in pressure)~\cite{pinaki,Harrison}.
The superconducting order parameter $\Delta_{sc}$ is self-consistently 
computed inside DMFT from the 
off-diagonal element of the matrix Green's function and it is 
found that for superconductivity the inter-orbital pairing amplitude 
is negligible due to the difference in their band energies, so I will be 
considering only the intra-orbital pairing amplitude. 
The total Hamiltonian now ($H=H_{n}+H_{res}^{(HF)}$) will be solved
 within multi orbital iterated perturbation theory (MOIPT)+ DMFT 
following earlier approaches~\cite{laad,ciuchi,attise2,aipp,garg}. 

\section{Dynamical Mean Field Theory}
\noindent To explore superconducting regime of 1T-TiSe$_2$, 
I used the MOIPT+DMFT with pressure dependent DOS. 
The anomalous Green's function:
$F(k,\tau) \equiv -\langle T_{\tau}c_{k\uparrow}(\tau)c_{-k\downarrow}$
$(0)\rangle$ 
satisfying $F(-k,-\tau)=F(k,\tau)$ for s wave pairing is introduced in the 
low temperature and high pressure phase of 1T-TiSe$_2$. 
Now in the presence of superconducting pairing the one-particle Green's function
 is 
\begin{eqnarray}
\hat{G}(k,\tau) \equiv -\langle T_{\tau} \Psi(k,\tau) \Psi^{\dagger}(k,0)\rangle \nonumber \\
\mbox{~~~~~~~~}= \left( \begin{array} {cc} G(k,\tau) & F(k,\tau) \\
F^{\dagger}(k,\tau) & -G(-k,-\tau) \end{array} \right)
\end{eqnarray}

and all the interactions are described through self energy matrix,
\begin{equation}
\hat{\Sigma}(k,i\omega_{n})=\left(
\begin{array}{cc} \Sigma(k,i\omega_{n}) & S(k,i\omega_{n}) \\
S^{\star}(k,-i\omega_{n}) & -\Sigma^{\star}(k,i\omega_{n})
\end{array}\right) \end{equation} where $\omega_{n}=(2n+1)\pi/\beta$ are
fermionic Matsubara frequencies and $S(i\omega_{n})$ contains pairing 
information.

\noindent Within the infinite dimensional symmetry formalism the k-dependence of 
anomalous Green's function will be through $\epsilon_k$ and the self energy 
is purely k independent \cite{kotliar}, so ${\Sigma} =
{\Sigma}(i\omega_{n})$ . Furthermore, since the SC order parameter
 is considered real in this system, then
$S(i\omega_{n})=S^{\star}(-i\omega_{n})$. 
Now the full lattice Green's function can be calculated using Dyson's 
equation as 
\begin{equation} 
\hat{G}^{-1}(k,i\omega_{n}) = \left(  \begin{array}{cc}
i\omega_{n}+\mu-\epsilon_{k} & 0
\\ 0 & i\omega_{n}-\mu+\epsilon_{k} \end{array}\right) - \hat{\Sigma}(i\omega_{n})
\end{equation}
$$ = \left(  \begin{array}{cc} i\omega_{n}+\mu-\epsilon_{k}-\Sigma(i\omega_{n})
& -S(i\omega_{n}) \\ -S(i\omega_{n}) &
i\omega_{n}-\mu+\epsilon_{k}+ \Sigma(-i\omega_{n})
\end{array}\right) $$

\noindent The self-consistency is obtained by putting the condition that the 
impurity Green's function coincides with the onsite Green's function of the 
lattice.

\noindent Since in TiSe$_2$ the normal state and the CDW ordered state is having major 
dependence on Ti-d and Se-p state and preformed excitons drive the incoherent 
normal state to coherent CDW state I will now extend the Green's function 
here to take the inter-orbital hopping and Coulomb interaction into account and 
will put orbital indices on the Green's function and self energy.

\noindent Following the earlier formalism \cite{kotliar,laad} 
to take inter-orbital hopping in Green's function I will now rewrite the full 
orbital dependent Green's function as 
$$
\hat{G_\nu}^{-1}(k,i\omega_{n}) = \left(  \begin{array}{cc} G_{\nu\nu}^{-1}
& -S(i\omega_{n}) \\ -S(i\omega_{n}) &
-G^{-1\star}_{\nu\nu}
\end{array}\right) $$
where
$$G_{\nu\nu}^{-1}=i\omega_{n}+\mu-\epsilon_{k\nu}-\Sigma_{\nu}(i\omega_{n})-\f{(t_{\nu\nu ^{'}}-\Sigma_{\nu\nu ^{'}})^{2}}{(i\omega_{n}+\mu-\epsilon_{k\nu ^{'}}-\Sigma_{\nu ^{'}}(i\omega_{n}))}$$

\noindent  Where $\nu(\nu ^{'}$ stands for a(b),b(a) bands. In the above equation $\Sigma_{\nu}, \Sigma_{\nu ^{'}}$, and $\Sigma_{\nu\nu ^{'}}$ 
are calculated  following earlier procedure \cite{attise2}. The electron lattice
 interaction and inter-orbital coulomb interaction contributes in orbital 
dependent self energy. Now to solve the above mentioned model 
within DMFT I have redesigned the IPT which is already an 
established technique for the repulsive Hubbard model in paramagnetic phase
\cite{kotliar,pruschke}. Though it is an approximate method it gives good 
qualitative arguement with the more exact methods to solve repulsive 
Hubbard model \cite{bulla,attise2,laad}. 
Also IPT has been successfully applied \cite{garg} in single band attractive 
Hubbard model. 
Reasonably this can be used in present context also.
The IPT technique for superconductivity is constructed in a way that it should 
successfully reproduce the self energy leading order terms 
in both the weak coupling limit 
and large frequency limit and it must be exact in the atomic limit. 
Considering all this the IPT self energy is computed for superconductivity. 
\noindent In the atomic limit, the
second order self energy vanishes as discussed earlier~\cite{garg}. 
Within this modified IPT approximation I have solved DMFT equation taking
$U_{aa}=U_{bb}=U$=0.5 eV and $U_{ab}$=0.1 (This values were calculated 
reproducing the normal state and CDW state) and the order parameter is computed 
from the off diagonal element of the matrix Green's function in presence of an 
infinitesimal symmetry breaking field. 

\begin{figure}
(a)
{\includegraphics[angle=270,width=0.8\columnwidth]{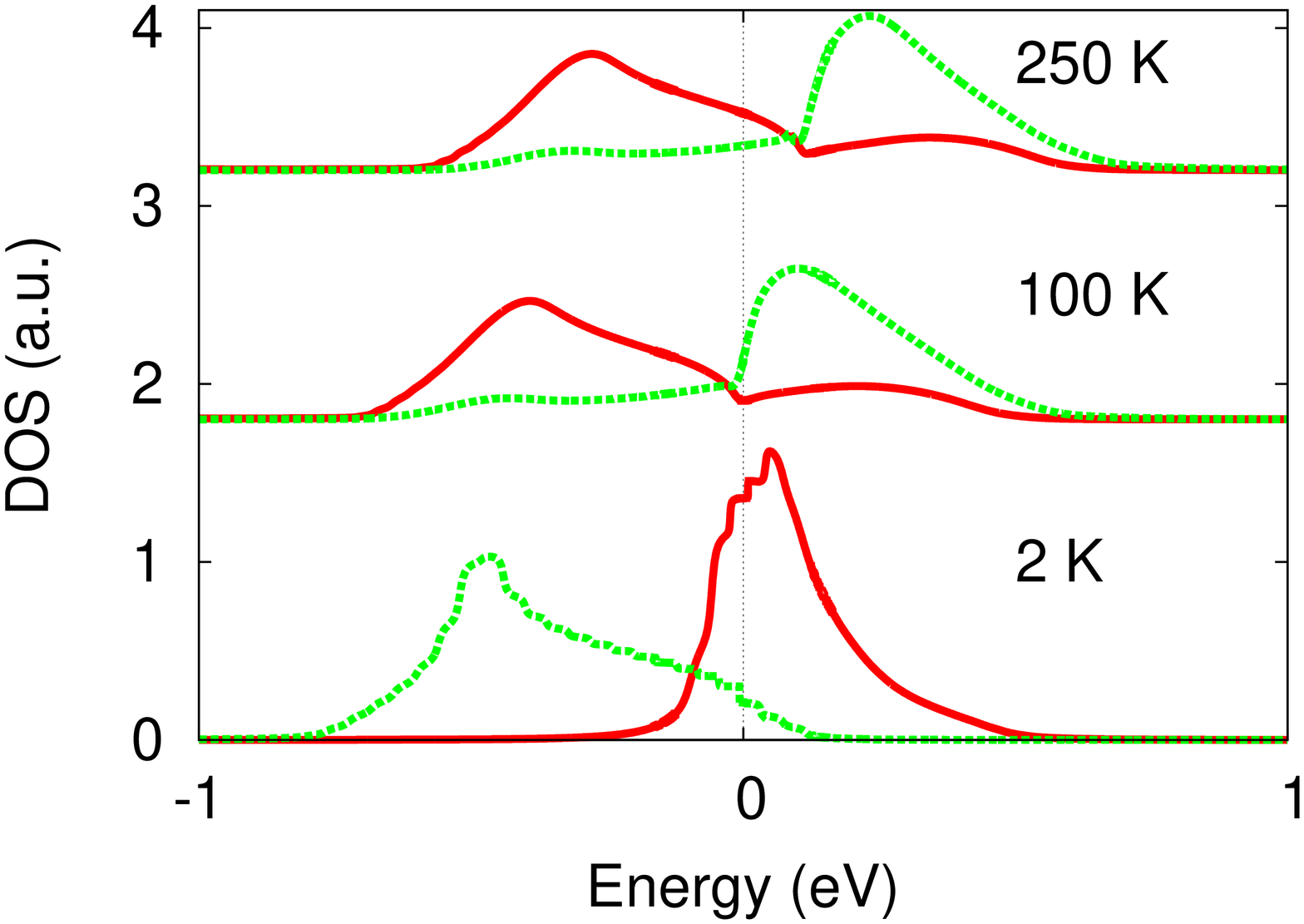}}

(b)
{\includegraphics[angle=270,width=0.8\columnwidth]{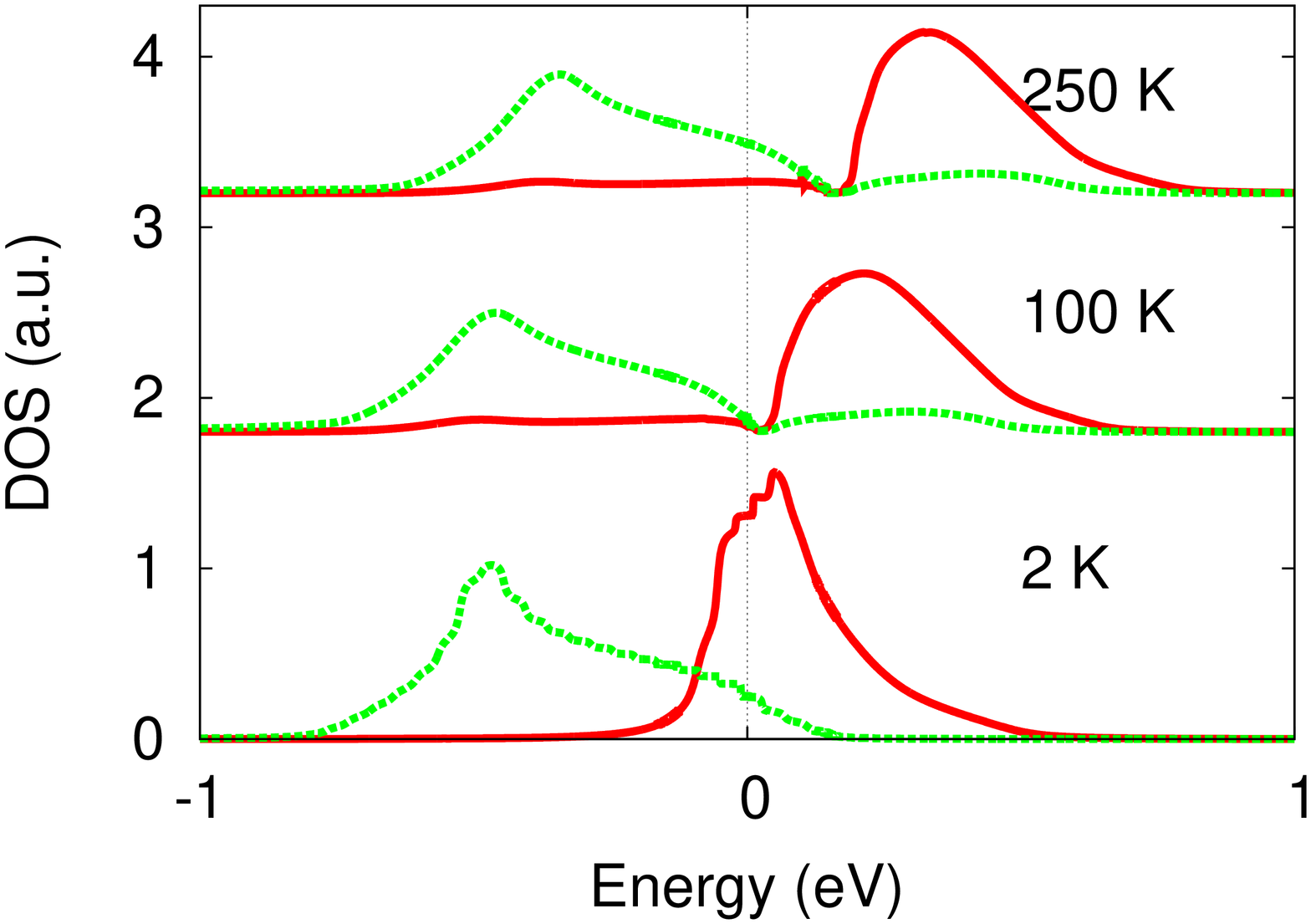}}
\caption{(Color Online) Density of states at three different temperature and two pressure, (a) 1.1 GPa and (b)2.3 GPa. This two different pressure 
at lower temperature represents CDW state and superconducting state. Red color stands for DOS of `a' band and green color stands for `b' band. DOS for different temperatures are shifted in the y-axis.}
\label{fig1}
\end{figure}
\begin{figure}
\centering
{\includegraphics[angle=270,width=0.8\columnwidth]{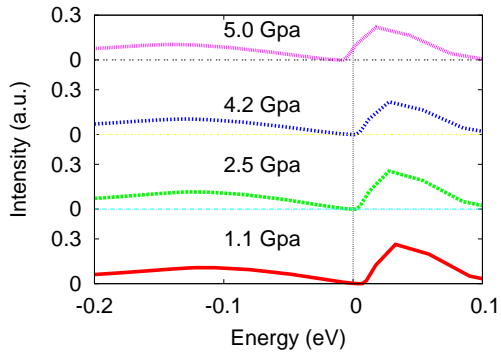}}
\caption{(Color Online) Theoretical ARPES spectra at `M' point in different 
pressure at 2 K temperature. As pressure increases from CDW state to normal 
state crossing a 
superconducting region the spectral density at Fermi level changes. A gap can be detected at the Fermi level at 2.5 and 4.2 GPa i.e. in the superconducting state.}
\label{fig2}
\end{figure}
\begin{figure}
\centering
{\includegraphics[angle=270,width=0.8\columnwidth]{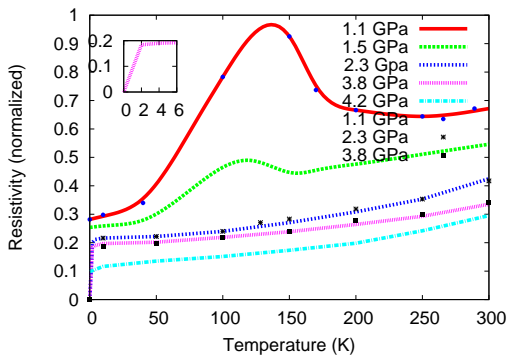}}
\caption{(Color Online) Temperature dependent DMFT resistivity at different 
pressure. DMFT results agree well with previous experimental results (represented by three types of colored points for three different pressure after A.F. Kusmartseva, et al.\cite{sipos}). The 
inset shows the resistivity at low temperature at 3.8 GPa pressure in the superconducting phase. Resistivity numerical data is normalized with respect to maximum value}
\label{fig3}
\end{figure}

\section{Result and Discussion}
\noindent I will show how the new DMFT approach described above 
explains a range 
of experimental data. DMFT renormalizes the LDA band position 
depending on its occupation by the intra and inter-orbital Hartree terms 
in self energy and dynamical correlations generically transfer spectral weight 
in larger energy scale. DMFT many body density of states (DOS) for Ti-d and Se-p
 band is shown at two different pressure in Fig.\ref{fig1} at three different 
temperatures. These three temperatures correspond to predicted normal state (250 K), CDW state (100 K),
\cite{attise2} and superconducting state (2 K). 
A clear `orbital selective' gap is found in the CDW state spectral function at 
both 1.1 GPa and 2.3 GPa but in the superconducting state the number density
 in the Fermi level increases significantly. Also at 250 K temperature the 
orbital selectivity vanishes. Interestingly with increasing pressure 
orbital selective 
pseudogap of the CDW phase increases as well as in the SC phase the number density 
at the Fermi level increases. The increasing number density at the Fermi level 
also supports that not only increased pressure but also doping the system will 
help in superconductivity. Surprisingly the spectral functions are not 
showing any conventional energy gap in the Fermi level. In 1T-TiSe$_2$ SC phase 
is following an unconventional CDW state and an excitonic normal state. So 
superconductivity here can also be associated with unusual gap. In this 
scenario momentum-resolved one particle spectral function can prove if there is 
presence of any energy gap in the Fermi surface.\\
\noindent In Fig.\ref{fig2} I show DMFT one-particle spectral functions,
 $A_{a,b}({\bf k},\omega)=-$Im$G_{a,b}({\bf k},\omega)/\pi$ at M point at 
different pressure. Absence of infrared Landau Fermi liquid quasiparticle poles 
indicates existence of a incoherent excitonic features in $E_{k,a,b}$ which is 
associated with a pseudogap in angle resolved photoemission spectra (ARPES) 
lineshapes in CDW state. Here I have chosen `M' point only because in the CDW 
state a `quasi-particle peak' (originated due to electron phonon coupling) is 
found experimentally and theoretically \cite{attise2,monney} at this point. 
Below the 
superconducting transition temperature a clear `gap' can be discernible in 
ARPES lineshapes at M point at both 2.3 GPa and 4.2 GPa pressure. This 
superconducting `gap' closes with increasing pressure as 1T-TiSe$_2$ goes 
through a transition into a metallic state. Existence of such type of 
superconducting gap at M point was also found earlier in doped 1T-Tise$_2$ \cite{qian}.     

\begin{figure}
\centering
{\includegraphics[angle=270,width=0.8\columnwidth]{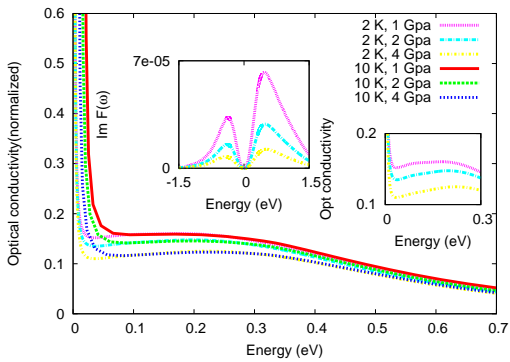}}
\caption{(Color Online) Optical conductivity (normalized with respect to the highest value) at three different pressure and at
two temperature for 1T-TiSe$_2$. The same (lower inset) at low energy presenting extended shoulder like feature appearing due to superconducting gap. Upper inset shows corresponding off diagonal Green's function at 2 K temperature.}
\label{fig4}
\end{figure}

\noindent Next I have computed transport properties of the 
dichalcogenide in high 
pressure. DMFT resistivity here is calculated from dc conductivity. Remarkably the DMFT resistivity (Fig.\ref{fig3}) at different pressure
 shows a very good accord with the earlier experimental resistivity data \cite{sipos}. 
At low pressure region upto 1.1 GPa the resistivity curve resembles with 
the curve of ambient pressure. The CDW peak is prominent and resistivity 
increases from normal state to CDW state and decreases afterwards. The strong 
peak in the resistivity curve signaling the CDW transition becomes gradually 
weak with the increase in pressure, and the maxima in -$d\rho(T)/dT$ also 
moves to lower temperature. In the low pressure phase resistivity above 
transition temperature shows a semimetallic behavior and the same is suspected 
well below the transition temperature\cite{wang}. Further with increase 
in pressure in the high temperature region 1T-TiSe$_2$ undergoes metallic 
behavior where 
resistivity manifests almost linear dependence with temperature in 
full qualitative accord with 
experiment\cite{sipos}. Also with increasing pressure the CDW peak in 
resistivity decreases and is destroyed completely above 2 GPa pressure and 
superconductivity is observed at about 2 K temperature. This excellently 
describes experimental resistivity data \cite{sipos} at different pressure (Fig.\ref{fig3}). In the inset of (Fig.\ref{fig3}) the low temperature part of 
resistivity at 3.8 Gpa pressure is shown in an extended frame. 
\\
In Fig.\ref{fig4} I show the normalized optical conductivity as a function of 
temperature and pressure. An extended shoulder like (right inset of 
Fig.\ref{fig4}) feature is observed at low temperature
 in the optical conductivity which reduces with increasing temperature 
to 6 K in the CDW state. This shoulder like feature is due to the 
superconducting gap found in DMFT ARPES spectra.  
Though there is still some sign of gap in optical conductivity 
at 1 GPa pressure but the off diagonal Green's functions shown in the inset 
display a comparatively larger integral than the superconducting ones which 
concludes that the gap is due to the CDW fluctuations. 
Optical conductivity is shown here till 0.7 eV without losing any
 important feature because I have considered only two band closer to the FL.
Presence of superconducting gap in the spectral function and observation of 
sharpness of the low energy optical conductivity is highly contrary to each 
other: it is not a Drude peak because at lowest temperature also the compound 
shows orbital selective non Fermi liquid behavior in DMFT. Moreover it is due to the coherence 
which sets 
in due to the superconducting order. If I check the off-diagonal Green's function 
at three different pressure at 2 K which also gives superconducting order 
parameter, the decrease in both the peaks is perceptible with increasing 
pressure. This detectable change in order parameter confirms the presence of 
phase transition  at particular pressure and temperature.      

\begin{figure}
(a)
{\includegraphics[angle=270,width=0.8\columnwidth]{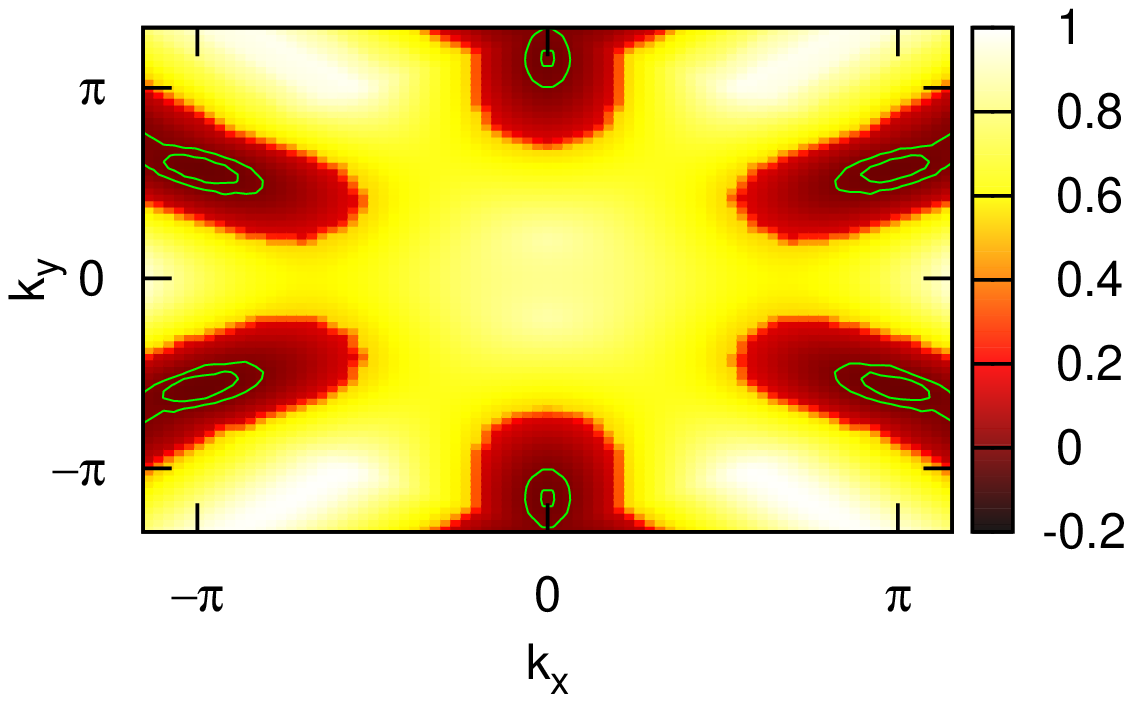}}

(b)
{\includegraphics[angle=270,width=0.8\columnwidth]{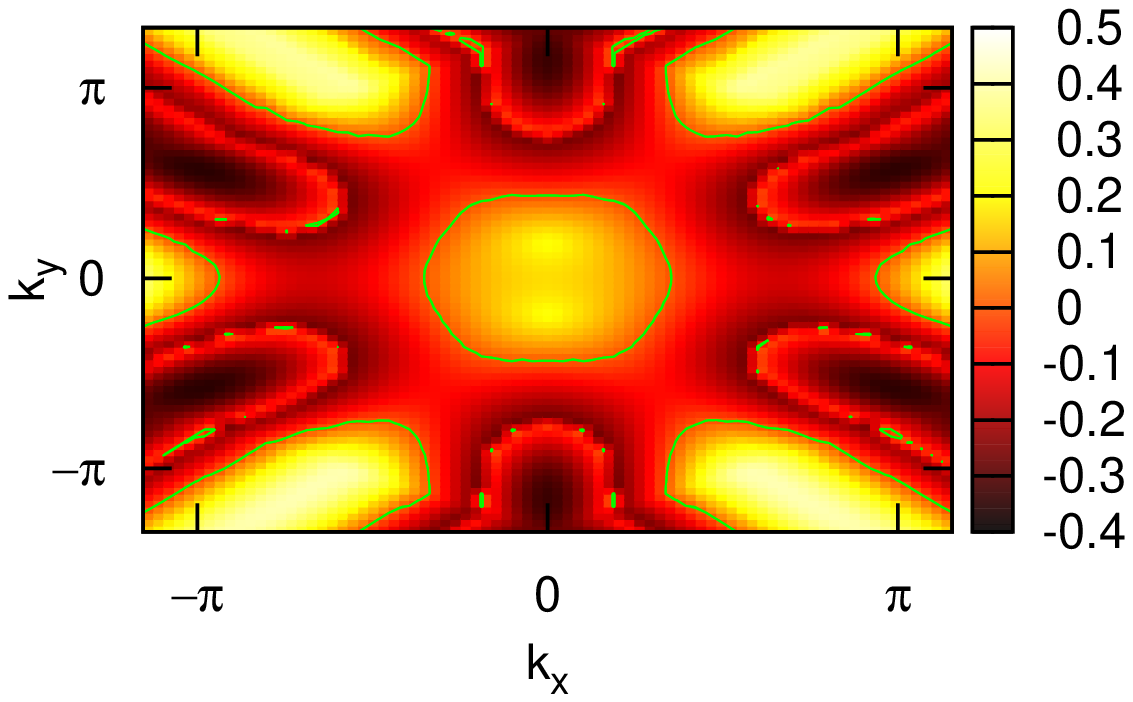}}
\centering

(c)
{\includegraphics[angle=270,width=0.8\columnwidth]{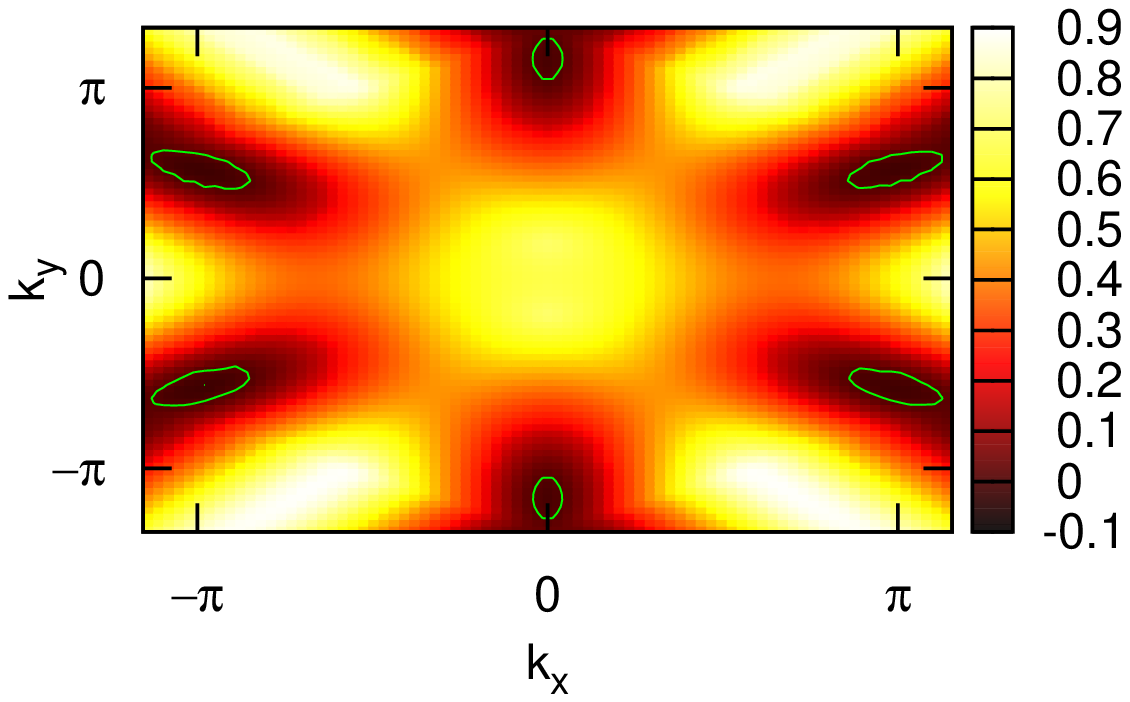}}
\caption{(Color Online) DMFT Fermi surface map in the (a)superconducting, (b)CDW and (c)normal state.}
\label{fig5}
\end{figure}
\begin{figure}
\centering
{\includegraphics[angle=270,width=0.8\columnwidth]{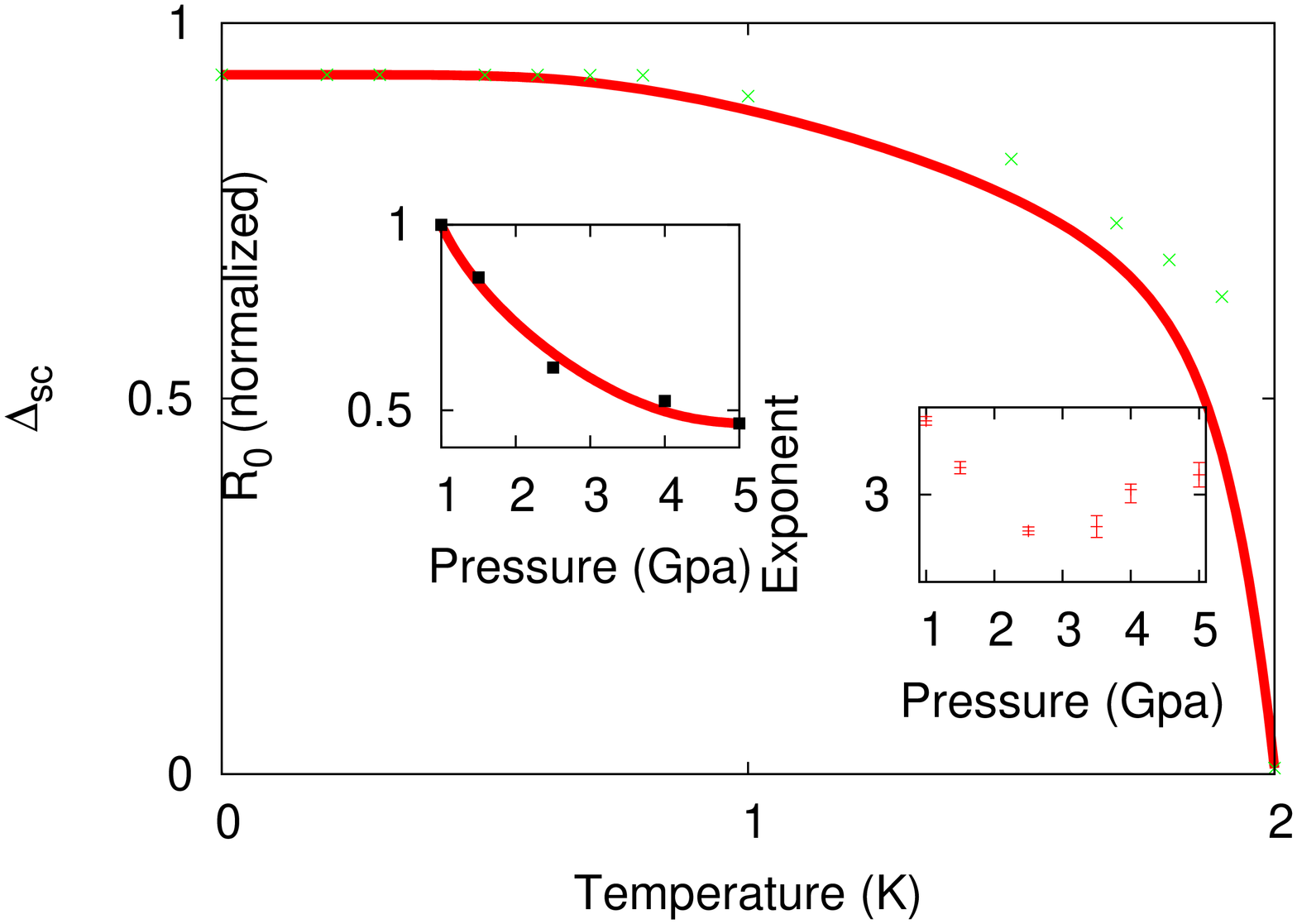}}
\caption{(Color Online) Temperature dependency of superconducting order 
parameter in the main panel. The left inset shows the residual resistivity 
(normalized with respect to the highest value) at 
different pressure. The residual resistivity shows good agreement with the previous experimental result (represented by black points after A.F. Kusmartseva, et al.\cite{sipos}). The right inset shows the 
resistivity exponent at different pressure.}
\label{fig6}
\end{figure}

\noindent DMFT Fermi surface (FS) map at different temperature is given in Fig.\ref{fig5}. 
FS in 1T-TiSe$_2$ does not rebuild across transition from normal state to CDW 
state in 1T-TiSe$_{2}$. This was one of the pillar to suggest 
existence of a dynamically 
fluctuating excitonic liquid at high temperature which is being coherent 
at low temperature i.e. the system undergoes transition into CDW state. 
But still enough information can be collected from FS in transition from CDW 
to superconducting phase. 
The band pockets spread out at normal state and, more
importantly, a brighter ring structure appears at the M point in the CDW state.
Now below superconducting transition temperature 
a much brighter elliptic structure in the middle of the hexagonal arms can be 
observed. Since these are found in the superconducting state so
 it may be a direct consequence of phase transition. Further the FS also shows 
isotropic s-wave superconducting gap. 
Moreover in the superconducting state there are electronlike pockets 
which develop from the sinking of band below FL. In case of doped samples of 
the same similar structure has been found in FS~\cite{qian}.
Now if this theory of superconductivity in this TMD is to be right then the 
superconducting order parameter which is determined self 
consistently within DMFT has to follow its nature of $(1-T/T_c)^{1/2}$. 
In Fig.\ref{fig6} main panel I have shown temperature dependent superconducting 
 order parameter which is calculated self consistently from DMFT: $\Delta_{sc} \propto \langle c_{a\downarrow}c_{a\uparrow}\rangle =-\f{|p|}{\pi} \int_{-\infty}^{0}
\mbox{Im~} F(\omega) d \omega $. The order parameter is following its nature 
and becomes almost zero near 2K, i.e at 
the transition temperature. Thus such quantitative argument built upon a 
novel PEL view between DMFT results and previous theoretical and experimental 
results support strongly the idea of a
dynamically fluctuating excitonic liquid at high $T$ giving way to a 
low-$T$ superconducting ordered state. \\
However, to be a dependable theory the {\it same} 
formulation must also comprehensively describe 
transport as well. In DMFT, this task is simplified: it is 
an excellent approximation to compute transport co-efficients directly from 
the DMFT propagators $G_{a,b}(k,\omega)$~\cite{silke}, since
(irreducible) vertex corrections rigorously vanish for one-band models, and
turn out to be surprisingly small even for the multi-band case. In Fig.
\ref{fig3} resistivity curve at different pressure is shown  which matches 
well with previous experiment both qualitatively and quantitatively. Furthermore 
I have also fitted the resistivity curves from 3K, above superconducting 
transition, to 30 K well below T$_{CDW}$ to find out the actual procedure from 
CDW to superconducting transition. Until 4 GPa pressure the residual resistivity
 (see inset of Fig.\ref{fig6}) manifests a large change.
Whereas only some 
marginal change in residual resistivity is found after superconducting phase. 
Detail study of the resistivity 
temperature exponent n, calculated from fitting the resistivity 
with $R(T)=R_0+AT^n$, shows also huge change across the pressure range where 
superconductivity appears and at the pressure where transition temperature is 
maximum in the {\it dome} this exponent `n' shows local minimum. 
Outside the 
superconducting dome the value of `n' remains about 3.0, 
which is in stark conflict with the idea of electron-electron or 
electron-phonon scattering which gives value of n as 2 or 5 respectively. This 
fact is rather unusual but the same is found in experiment also~\cite{sipos}.
Wilson \cite{wilsona} proposed this higher power law temperature dependence 
of the resistivity as scattering from low density band to a higher one. 
Though at the superconducting pressure region this exponent dovers around a 
value of 2.6 which represents a quantum critical scenario and quantum 
fluctuation around that region.\\
In the LDA bands of TiSe$_2$ there is negative indirect band gap. Increased pressure or doping closes the gap and 
some of the low energy levels cross the Fermi energy. 
Thus with increasing pressure CDW order is also destroyed. 
Increasing the number density at the FL 
enhances superconducting pairing correlations in a system. It has been 
observed that with increasing pressure the number density at the FL (D(E$_f$))
increases. Such an increase in D(E$_f$) in turn increases 
superconducting  transition temperature T$_c$ in many 
systems as expected also from the formula $K_BT_c=\hbar \omega exp{-1/[D(E_f)V]}$.
Whereas increase in pressure beyond 4 GPa closes the superconducting 
gap found at M point
and the dichalcogenide behaves like a metal.

\vspace{0.5cm}

\section{Conclusion}
\noindent In the light of these DMFT results, it can be concluded that with pressure 
the preformed excitons in the normal 
state drives the compound to undergo a CDW superconducting phase transition. 
Though conventional s-wave gap is absent in the density of states but ARPES 
spectra shows the presence of superconducting gap at $M$ point, which is 
confirmed in FS plots also. An electronlike pocket is found to grow with 
applied pressure in the superconducting phase of this dichalcogenide. My 
approach highlights the role of pressure associated with excitonic correlation 
in small carrier density systems. Based upon a combination of high pressure 
LDA bandstructure and a DMFT treatment of dynamical interband excitons coupled 
to phonons, I have shown that this calulation describes a wide variety of 
features in this system. I have presented both qualitative and quantitative 
scenario for the emergence of unconventional superconductivity as a direct 
outcome of exciton phonon coupling at different pressure. 
Also in other parent or impurity doped TMDs superconductivity arises from 
normal states following a CDW order the present scenario should be  
relevant to all these cases.\\ 
\noindent SK acknowledges useful discussion and collaboration on similar 
systems with Arghya Taraphder.

\end{document}